\def\selectedoptions{final}

\documentclass[\selectedoptions]{aipproc}
\usepackage{amsmath}
\usepackage{epstopdf}
\layoutstyle{8x11double}

\begin{document}

\title 
      [INFLUENCE OF CONTROLLED VISCOUS DISSIPATION ON THE PROPAGATION OF STRONGLY NONLINEAR WAVES IN STAINLESS STEEL BASED PHONONIC CRYSTALS]
      {INFLUENCE OF CONTROLLED VISCOUS DISSIPATION ON THE \\PROPAGATION OF STRONGLY NONLINEAR WAVES IN \\STAINLESS STEEL BASED PHONONIC CRYSTALS}


\author{Eric B. Herbold}{
  address={Department of Mechanical and Aerospace Engineering, University of California at San Diego, La Jolla, California 92093-0411, USA},
  email={eherbold@ucsd.edu},
  thanks={}
}

\iftrue
\author{Vitali F. Nesterenko$^{*}$}{
  address={Materials Science and Engineering Program, University of California at San Diego, La Jolla, California 92093-0418, USA},
}
\author{Chiara Daraio}{
  address={Materials Science and Engineering Program, University of California at San Diego, La Jolla, California 92093-0418, USA},
}
\fi

\copyrightyear  {2005}

\begin{abstract}
Strongly nonlinear phononic crystals were assembled from stainless steel spheres.  Single solitary waves and splitting of an initial pulse into a train of solitary waves were investigated in different viscous media using motor oil and non-aqueous glycerol to introduce a controlled viscous dissipation.  Experimental results indicate that the  presence of a viscous fluid dramatically altered the splitting of the initial pulse into a train of solitary waves.  Numerical simulations qualitatively describe the observed phenomena only when a dissipative term based on the relative velocity between particles is introduced.\\
\end{abstract}

\date{\today}

\maketitle

\subsection{INTRODUCTION}

In recent years, many experimental and numerical efforts have been focusing on the propagation of strongly nonlinear solitary waves in granular media \cite{VFN_01}.  These waves are a natural extension of the well known weakly nonlinear solitary waves such as the Korteweg-de Vries (KdV) solitons.  Chains composed of steel, brass, glass, Nylon \cite{VFN_01, CFF_97, CG_98} and Polytetraflouroethylene (PTFE) \cite{DNHJ_05} particles support solitary waves using Hertz' law for particle interactions. 

However, dissipation plays a significant role in the transmission of pulses in almost all experimental settings.  Restitution coefficients and velocity dependent friction have been used to investigate dissipation in a chain of particles \cite{MSH_01, RL_03}.  The energy losses are significant even when the experiments are performed in air (see Fig. 1.25 in \cite{VFN_01}).  These losses may be attributed to uncontrolled features in the experimental setup as well as inherent material properties \cite{JMSS_04}.

To predictably control the dissipation, a chain of spheres may be immersed in viscous fluids.  The viscous dissipation of pulses in chains of particles differs from a two particle interaction in liquid because the compression wave dominates the system's dynamic behavior.  To our knowledge, there has not been an attempt to extend the research involving the  collision of particles in fluids to a chain of particles.  This paper extends current knowledge of  a two particle collision in a viscous medium to a case where multiple particle interactions support a solitary wave.

\subsection{ANALYTICAL MODEL AND NUMERICAL CALCULATIONS}

The analytical model for a one-dimensional Hertzian chain of spheres can be found in \cite{VFN_01}.  The description of particle interactions in fluids is presented in [8-12].  One of the complications of using current sphere-fluid models for a chain of spheres is that a developed flow around the sphere is assumed before the particle collision though the duration of the pulse in a chain is relatively short ($\sim$50 $\mu$s).  Nevertheless we use the outlined approach for the first stage of our research.  In \cite{M-T_68} two particles are considered to be traveling towards each other in a surrounding medium.  If the spheres are the same size then the total kinetic energy is
\begin{equation}
\label{eq:MT}
\frac{1}{4}\left(2m + m' + \frac{3}{8}\frac{m'R^{3}}{h^{3}}\right)U^{2} = constant,
\end{equation}
where $m = \frac{4}{3}\pi\rho R^{3}$ is the mass of the particle, $m' = \frac{11}{12}\pi \rho_{f}R^{3}$ is the added mass, $R$ is the particle's radius, $h$ is the separation distance between particle centers and $U=\frac{dh}{dt}$ is the relative particle velocity.  Differentiating Eq. (\ref{eq:MT}) with respect to $h$ yields the added mass and pressure force terms. The equations of motion for a chain of spheres, with mass $m_{i}$, placed vertically in a gravitational field in a fluid becomes
\begin{multline}
\label{eq:full}
\left(m_{i}+m'_{i}\right)\frac{d^{2}x_{i}}{dt^{2}} = F_{c,i}\left(\delta\right) + m_{i}g + F_{b,i} \\+ F_{D,i}
+ F_{p,i}  + F_{d,i}.  
\end{multline}
where $\delta_{i,i+1}=\left(R_{i}+R_{i+1}+x_{i}-x_{i+1}\right)$ and the $x_{i}$'s are the particle positions.  The compression force $F_{c,i}$ is based on Hertz law between particle '{\it{i}}' and the adjacent particles '{\it{i}}-1' and '{\it{i}}+1';
\begin{equation}
F_{c,i}  = \phi\left(\delta_{i-1,i}\right)- \psi\left(\delta_{i,i+1}\right),
\end{equation}
where $\phi$ and $\psi$ can be expressed as
\begin{equation}
\label{eq:phi}
\phi\left(\delta_{i-1,i}\right) =A_{i-1,i}\left(\delta_{i-1,i}\right)^{3/2}
\end{equation}
and
\begin{equation}
\label{eq:psi}
\psi\left(\delta_{i,i+1}\right)=A_{i,i+1}\left(\delta_{i,i+1}\right)^{3/2}.
\end{equation}
The coefficients in Eqs. (\ref{eq:phi}) and (\ref{eq:psi}) are identical except for a shift of indices.  The equation for $A_{i-1,i}$ can be inferred from 
\begin{equation}
A_{i,i+1} = \frac{4E_{i}E_{i+1}\left(\frac{R_{i}R_{i+1}}{R_{i}+R_{i+1}}\right) ^{1/2}}{3\left[E_{i+1}\left(1-\nu_{i}^{2}\right) + E_{i}\left(1- \nu_{i+1}^{2}\right)\right]},
\end{equation}
where $E_{i}$ and $\nu_{i}$ are the elastic modulus and Poisson's ratio of the particle.
For all of the calculations performed with air as the surrounding medium, only the first two terms on the right hand side and the first term on the left side of Eq. (\ref{eq:full}) are used.  When liquids surround the chain of spheres the buoyancy, drag, pressure, added mass and dissipative terms are used.  The buoyancy force for each particle is
\begin{equation}
\label{eq:buo}
F_{b,i}= -\frac{4}{3}\pi R_{i}^{3}\rho_{f}g,
\end{equation}
where $g$ is the gravitational constant, $R_{i}$ is the radius of the particle and  $\rho_{f}$ is the density of the fluid.
The drag force has a correction factor to account for a Reynold's number $Re$ greater than unity i.e. not in the Stoke's flow regime,  
\begin{equation}
\label{eq:drg}
F_{D,i} = -6\pi \nu R_{i}U_{i}(1+0.15Re^{0.687}),
\end{equation}
where $\nu$ is the dynamic viscosity and $U_{i}$ is the particle velocity.
The pressure force in \cite{ZFZP_99} is written for one sphere moving toward a stationary sphere.  In our case, we use a relative velocity between particles,
\begin{equation}
\label{eq:pres}
F_{p,i} = \frac{3}{8}\pi R_{i}^{2}\rho_{f}\left[(U_{i-1}-U_{i})^{2}-(U_{i}-U_{i+1})^{2}\right].
\end{equation}

We introduce an additional dissipative term based on the relative velocity between particles with a fitting parameter $c$ due to the lack of a qualitative agreement between the experiments and calculations based on the drag force term (Eq. (\ref{eq:drg})). This addition is tantamount to adding a dash-pot between neighboring particles \cite{DML_69},
\begin{equation}
\label{eq:diss}
F_{d,i} = c\left(U_{i-1} - 2U_{i}+U_{i+1}\right),
\end{equation}
where the coefficient $c$ is a fitting parameter.  The physical reason for this term can be due to the radial flow of liquid caused by the change of contact area between particles.

MATLAB's intrinsic ODE45 solver was used to march the explicit calculation forward in time with a time-step of 0.05$\mu$s.  The error in the energy calculations were found to be less than 10$^{-9}$\% in air and within 10$^{-5}$\% in the fluid.  The error in the conservation of linear momentum performed with a chain in air was less than 10$^{-12}$\%.

\subsection{EXPERIMENTAL PROCEDURES AND RESULTS}

In the experiments, the impulse propagation was investigated in three different media: air, SAE 10W-30 motor oil, and non-aqueous Glycerol GX0185-5.  
\begin{figure*}[t]
\label{fig:1}
\includegraphics[scale=0.41,clip,trim=14mm 11mm 16mm 15mm]{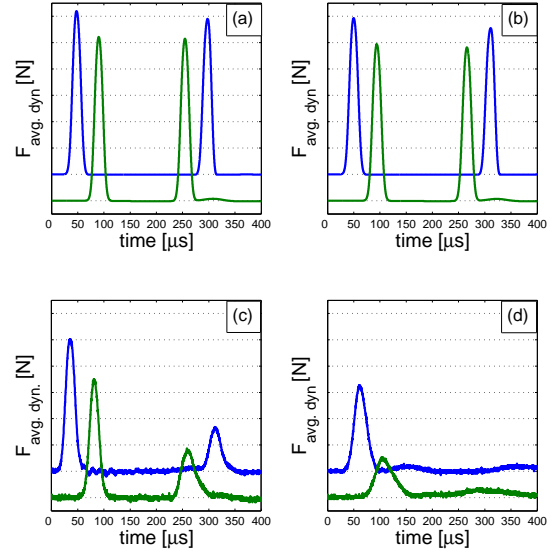}
\caption{Single solitary wave in calculations (a) and experiments (c) in a chain surrounded by air. Results for an identical chain surrounded by glycerol in calculations (b) and experiments (d).  Vertical scale is 2 N/div.}
\end{figure*}
The density and dynamic viscosity used in numerical calculations were $\rho_{f}=880$ kg/m$^{3}$, $\nu=0.067$ Ns/m$^{2}$ for oil and  $\rho_{f}=1260$ kg/m$^{3}$, $\nu=0.62$ Ns/m$^{2}$ for glycerol.  An experiment was performed with each of the three types of media to see how single and multiple solitary waves propagate through the chain.  The chain was placed into an adjustable holder that had four contact points on each sphere.  The air or fluid was able to flow freely between the contacts as opposed to the cylindrical holder that had been used in previous experiments.  To create a single solitary wave, a spherical stainless steel striker with a radius of $R=4.76$ mm and a mass of $m=0.4501$ g was used to impact the top of a chain of 19 stainless steel particles (also with $R=4.76$ mm) with a velocity of $U_{0}=0.44$ m/s.  To create multiple solitary waves in the same chain, a cylindrical alumina striker with a larger mass of $m=1.23$ g impacted the chain at $U_{0}=0.44$ m/s.  The elastic modulus and Poisson's ratio of the stainless steel particles were 193 MPa and 0.3, respectively.  The experimental results were recorded via piezoelectric gauges imbedded \cite{DNHJ_05} in the 10th and 15th particles from the top of the chain.  The procedures outlined in \cite{DNHJ_05} were implemented to compare the calculations to the experimental results using the averaged dynamic force as noted on the vertical axis of Fig. \ref{fig:1}, \ref{fig:2} and \ref{fig:3}.

In Fig. \ref{fig:1} the numerical and experimental results are shown for a single solitary wave in a chain surrounded by air (a), (c) and glycerol (b), (d).  It is evident that there is a very small difference between the numerical results for the chain in air and glycerol using Eqs. (\ref{eq:full})-(\ref{eq:pres}), which prompted the inclusion of the additional dissipative term (Eq. (\ref{eq:diss})).  Without this dissipative term, the numerical results for glycerol shown in Fig. \ref{fig:1}(b) do not exhibit the shape and speed of the pulses in experiments.

In experiments the speeds of the single pulse were $V_{s}=520$ m/s in air and  increased to $V_{s}=541$ m/s in glycerol.  It is interesting to note that the calculated signal speed was $V_{s}=564$ m/s in air and decreased to $V_{s}=540$ m/s in glycerol.  In numerical calculations the speed of the single pulses in chains surrounded by both air and glycerol should be higher than the experimental pulse speeds due to the higher amplitude of the waves.  Also, one would intuitively think that the pulse speed in air would be higher than in glycerol due to viscous dissipation.  This is the case in calculations (even when the additional dissipative term $F_{d}$ is added) but the opposite is true in experiments (Fig. \ref{fig:1}(c) and (d)).

\begin{figure*}[htbp]
\label{fig:2}
\includegraphics[scale=0.37,clip,trim=17mm 0mm 18mm 5mm]{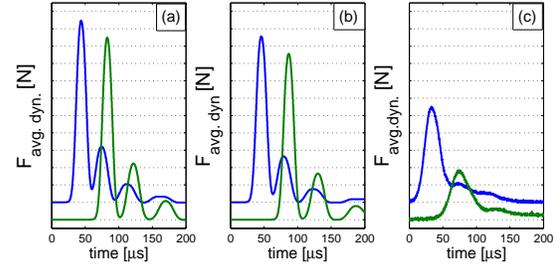}
\caption{(a) Numerical results for multiple solitary waves in a chain of 19 particles surrounded by air.  (b) Numerical results for multiple solitary waves in an identical chain surrounded by glycerol.  (c) Experimental results related to (b).  Vertical scale is 2 N/div.}
\end{figure*}

 In Fig. \ref{fig:2} the numerical results for a train of solitary waves in air and glycerol are presented.  The dissipative term $F_{d}$ was not included in the calculations in (a) and (b) and it is apparent that the amplitudes of the waves are much higher than the experimental results in glycerol (c).  Additionally, no signal splitting into a train of solitary waves were present in experiments Fig. \ref{fig:1}(c), which is not reflected by the calculations without the additional dissipative term $F_{d}$.  Again, the wave speeds in experiments were higher in glycerol than in air.
 
There is a disparity between the presented analytical formulation using the corrected Stokes drag for dissipation and the experiments.  The largest dissipative term in the calculations (before the addition of the dissipative term $F_{d}$) is the drag force term.  The reduction of the amplitude of a single pulse was about  3\% in oil and about 4\% in glycerol when using Eqs. (\ref{eq:full})-(\ref{eq:pres}).  These equations, while important for accurately describing a particle trajectory pre and post collision are negligible when their effect on the compression wave are examined.
 
 \begin{figure*}[tbp]
\label{fig:3}
\includegraphics[scale=0.34,clip,trim=19mm 0mm 20mm 9mm]{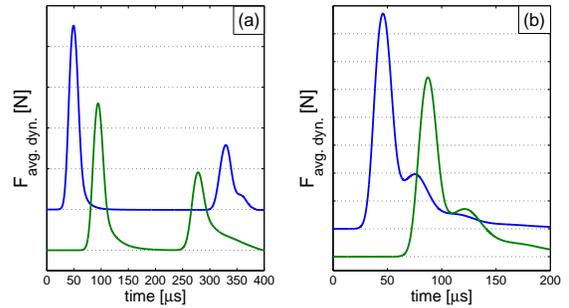}
\caption{(a) Numerical results for a single pulse with additional dissipative term Eq. (\ref{eq:diss}) in a chain surrounded by glycerol.  (b) Numerical result for an impact by an alumina striker on the same chain.  Vertical scale is 2 N/div.}
\end{figure*}
 
The results of adding the dissipative term Eq. (\ref{eq:diss}) are presented in Fig. \ref{fig:3} for single and multiple pulses traveling in a chain submersed in glycerol.  A similar asymmetry and widening of the pulse, apparent in experiments, can also be seen when Fig. \ref{fig:3}(a) is compared to Fig.\ref{fig:1}(d).  The lower line in Fig. \ref{fig:1}(d) has a noticeable shock-like tail.  In calculations the tail of the wave becomes more shock-like as parameter $c$ increases.  The qualitative behavior of the pulse in experiments is reproduced in calculations by adding an additional dissipative term $F_{d}$ with a coefficient $c=6.0$ Ns/m for glycerol and $c=0.648$ Ns/m for oil.  These values provided the best comparison between experimental and numerical data and were scaled according to the difference in viscosity.  When comparing Fig. \ref{fig:2}(b) and Fig. \ref{fig:3}(b), notice the additional dissipation Eq. (\ref{eq:diss}) has also created a shock-like response of the train of solitary waves.  In calculations the amplitude and tendency to split decreased by adding this term in accord with experiments.

\subsection{CONCLUSIONS}
The experimental results indicate a qualitative change of the propagating shock and solitary waves in a chain immersed in glycerol while only a small change in oil.  Without the relative velocity based dissipative term $F_{d}$, the equations pertaining to the surrounding fluid could not accurately reproduce the amplitude or the shock like structure of the incident pulse.  This term provided the qualitative change needed to match the numerical analysis to the experiments.  The numerical analysis predicted a decrease in solitary wave speed as the viscosity of the surrounding fluid increased contrary to experiments.  This phenomenon may be explained by an increased effective stiffness modulus between particles in the presence of a viscous fluid.
\subsection{ACKNOWLEDGEMENTS}
The  work is supported by NSF (Grant No. DCMS03013220).

\bibliographystyle{aipprocl}

\bibliography{waves_viscous_aps.bbl}

\begin{thebibliography}{10}
\providecommand{\enquote}[1]{``#1''}
\expandafter\ifx\csname url\endcsname\relax
  \def\url#1{\texttt{#1}}\fi
\expandafter\ifx\csname urlprefix\endcsname\relax\def\urlprefix{URL }\fi

\bibitem{VFN_01}
Nesterenko, V.~F., \emph{Dynamics of Heterogeneous Materials}, Chapter 1,
  Springer-Verlag, NY, 2001.

\bibitem{CFF_97}
Coste, C., Falcon, E., and Fauve, S., \emph{Physical Review E}, \textbf{56},
  6104 (1997).

\bibitem{CG_98}
Coste, C., and Gilles, B., \emph{The European Physical Journal B}, \textbf{72},
  155--168 (1999).

\bibitem{DNHJ_05}
Daraio, C., Nesterenko, V., Herbold, E., and Jin, S., \emph{Physical Review E},
  \textbf{72}, 016603 (2005).

\bibitem{MSH_01}
Manciu, M., Sen, S., and Hurd, A.~J., \emph{Physica D}, \textbf{157}, 226--240
  (2001).

\bibitem{RL_03}
Rosas, A., and Lindenberg, K., \emph{Physical Review E}, \textbf{68}, 041304
  (2003).

\bibitem{JMSS_04}
Job, S., Melo, F., Sen, S., and Sokolow, A., \emph{Physical Review Letters},
  \textbf{94}, 178002 (2005).

\bibitem{GLP_01}
Gondret, P., Lance, M., and Petit, L., \emph{Physics of Fluids}, \textbf{14},
  643--652 (2001).

\bibitem{ZFZP_99}
Zhang, J., Fan, L.-S., Zhu, C., Pfeffer, R., and Qi, D., \emph{Powder
  Technology}, \textbf{106}, 98--109 (1999).

\bibitem{M-T_68}
Milne-Thomson, L.~M., \emph{Theoretical Hydrodynamics}, Macmillan Education,
  London, 1968, 5th edn.

\bibitem{DSH_85}
Davis, R.~H., Serayssol, J.-M., and Hinch, E., \emph{Journal of Fluid
  Mechanics}, \textbf{163}, 479--497 (1985).

\bibitem{H72}
Hocking, L.~M., \emph{Journal of Engineering Mathematics}, \textbf{7}, 207--221
  (1972).

\bibitem{DML_69}
Duvall, G.~E., Manvi, R., and Lowell, S.~C., \emph{Journal of Applied Physics},
  \textbf{40}, 3771--3775 (1969).

\end{thebibliography}

\end{document}